\documentclass[aps,prb,twocolumn,groupedaddress,showpacs]{revtex4}

\usepackage{graphicx}

\begin{document}

\title{Effect of magnetism on the lattice dynamics in the $\sigma$-phase
$Fe-Cr$ alloys}

\author{S.M. Dubiel}
\email[Corresponding author: ]{stanislaw.dubiel@fis.agh.edu.pl}
\author{J. Cieslak}
\affiliation{
AGH University of Science and Technology,
Faculty of Physics and Applied Computer Science,
al. Mickiewicza 30, 30-059 Krakow, Poland
}

\author{M. Reissner}
\affiliation{%
Institute of Solid State Physics,
Viena University of Technology,
Wiednerhauptstr. 8-10, A-1040 Wien, Austria
}%

\begin{abstract}
Anomalies in the temperature dependences of the recoil-free factor, $f$, and
the average center shift, $\langle CS\rangle $, measured by $^{57}$Fe M\"ossbauer Spectroscopy, were
observed for the first time in the archetype of the $\sigma$-phase alloys
system, $Fe-Cr$.  In both cases the anomaly started at the temperature close
to the magnetic ordering temperature, and in both cases it was indicative of
lattice vibrations hardening.  As no magnetostrictive effects were found,
the anomalies seem to be entirely due to a spin-phonon coupling.  The
observed changes in $f$ and in $\langle CS\rangle $ were expressed in terms of the underlying
changes in the potential, $\Delta E_p$, and the kinetic energy, $\Delta E_k$, respectively.  The former,
with the maximum value larger by a factor of six than the latter, decreases,
while the latter increases with $T$.  The total mechanical energy change, $\Delta E$, was, in general, not
constant, as expected for the Debye-like vibrations, but it resembled that of
$\Delta E_p$.  Only in the range of 4-15 K, $\Delta E$ was hardly dependent on $T$.
\end{abstract}


\maketitle

A thorough knowledge and good understanding of atomic lattice vibrations in
solids, in general, and in technologically important materials, in
particular, is essential, for the proper understanding of their physical
properties such as thermal conductivity, heat capacity, vibrational entropy,
Debye temperature, electron-phonon coupling as well as the noise of
electronic devices.  One of the open questions in the field is a possible
relationship between magnetism and the lattice vibrations.  The contribution
of an electron-phonon interaction to magnetization of metallic systems is
expected to be small, as, in general, $\hbar \omega_D \over E_F$ is of
the $10^{-2}$ order \cite{Kim82},
where $\omega_D$ is the Debye cut-off frequency and $E_F$ is the Fermi energy.
Consequently, the effect of magnetism on the lattice dynamics in such systems should be rather
negligible.  However, following Kim \cite{Kim82} the effect of the electron-phonon
coupling can be strongly enhanced below the Curie temperature, $T_c$, in an
itinerant ferromagnet.  A good candidate for verifying these predictions seem
to be $\sigma-FeCr$ alloys which are believed to be itinerant ferromagnets
with $T_c$-values below 50 K \cite{Cieslak08}.

The $\sigma$-phase constitutes a broad class of binary and ternary alloy
systems with common crystallographic structure $(D^{14}_{4h} - P4_2/mnm)$ and
physical properties depending on the system \cite{Hall66}.  The $\sigma-FeCr$ which was
discovered in 1923 \cite{Bain23} and identified in 1954 \cite{Bergman54} has
been known not only as the archetype of $\sigma$-phase, but mainly for scientific and
technological reasons.  The former stems from its interesting physical
properties (e.  g.  complex crystallographic structure, high brittleness and
hardness, low-temperature magnetism and unusually high value of the specific
heat).  The latter follows from the deteriorating effect of the phase
presence on mechanical and corrosive properties of ferritic stainless steels
that are regarded as important construction materials.  Despite
the phase has been known for over a half of a century, only few papers
relevant to its lattice vibrational properties have been published so far
\cite{Cieslak02, Cieslak05, Dubiel10}.
The Fe-partial phonon density of states (PDOS) of the $\sigma-FeCr$
was found to be significantly different from that of the $\alpha-FeCr$,
which was not the case for the Debye temperatures of the two phases \cite{Dubiel10}.
In this Letter an evidence obtained from M\"ossbauer spectroscopic (MS) study is reported
 that magnetism can significantly affect the lattice dynamics in $\sigma$-phase Fe-Cr samples.

MS is an exceptionally well-suited tool to study the atomic vibrations in
solids as it delivers relevant information via two spectral
parameters viz.  the recoil-free factor, $f$, and the center shift, $CS$.
The former is related to the mean-square amplitude of vibrations of the
M\"ossbauer atoms, $\langle x^2\rangle $, through the following equation,

\begin{equation}
   f = exp(-\langle x^2\rangle k^2)
\label{e1}
\end{equation}

where $k$ is the wave vector of the gamma radiation.  The latter is a measure
of the mean-square velocity of the vibrating atoms, $\langle v^2\rangle $, via the
second-order Doppler shift, SOD, given by,

\begin{equation}
   SOD=-E_{\gamma}\langle v^2\rangle /2c^2
\label{e2}
\end{equation}

where $E_{\gamma}$ is the energy of the gamma rays.

In solids with no electron-phonon coupling, the temperature dependences of both
$CS$ and $f$ are smooth monotonic decreasing functions of $T$.  If, however,
a strong enough spin-phonon coupling is present, an anomaly in the two
quantities should be seen close to $T_c$.

\begin{figure*}[tp]
\includegraphics[width=.95\textwidth]{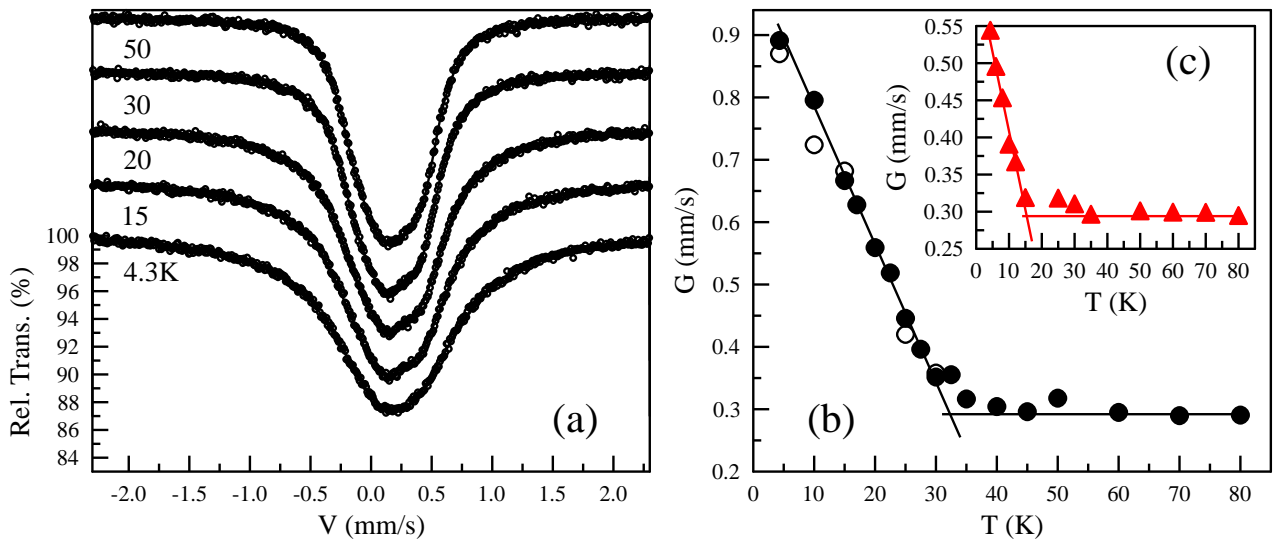}%
\caption{(color online)
(a) $^{57}$Fe M\"ossbauer spectra recorded on $\sigma-Fe_{54}Cr_{46}$ at various
temperatures shown in zero external magnetic field, and
the full line width at half maximum, G, versus temperature for (b)
$\sigma-Fe_{54}Cr_{46}$ and (c) $\sigma-Fe_{52}Cr_{48}$.  The intersection of the lines
defines the magnetic ordering temperature $T_c$. In (b) the data obtained for $\sigma-Fe_{54}Cr_{46}$
with a second run (open circles) are added.
 }
\label{F01}
\end{figure*}

Two samples of the $\sigma$-phase $FeCr$ alloys viz.  $Fe_{54}Cr_{46}$ and
$Fe_{52}Cr_{48}$ were the subject of the present study.  They were obtained
by an isothermal annealing of the bcc master alloys at $T=973$~K.  More
detailed description can be found elsewhere \cite{Cieslak08}.  In
search for a possible effect of magnetism on the atomic vibrations, $^{57}$Fe
M\"ossbauer spectra were recorded in a transmission geometry at various
temperatures, $T$, using a standard spectrometer, $^{57}$Co/Rh source for the
$\gamma$ 14.4 keV radiation and a flowing cryostat that enabled stabilization
of $T$ with an accuracy of $\pm 0.1$~K.  Examples of the spectra recorded in this
way can be seen in Fig.  1a.  They were analyzed using a least-square
iteration procedure.  Among the fitted parameters were those pertinent to the
present considerations viz.  the width of the Lorentzian-shaped line, $G$, as
well as the center shift, $CS$.  A more detailed description of the procedure
is given elsewhere \cite{Cieslak02}.  From the temperature dependence
of $G$, which can be seen in Figs.  1b,c, the value of the magnetic ordering
temperature (Curie point), $T_c$, was determined.  For $Fe_{54}Cr_{46}$ $T_c$
= 32.9 K and for $Fe_{52}Cr_{48}$ $T_c$ = 15.2 K were found.  These values
agree rather well with the corresponding figures of 38.9 K and 17.2 K,
respectively, found from the magnetization data \cite{Cieslak04}.

The temperature dependence of the average center shift,$\langle CS\rangle $, as found from the fitting procedure, is
illustrated in Fig.  2a.  A smooth change of $CS$ with $T$, as expected from
the Debye model, can be seen until $T$ reaches a certain critical value
below which a steep decrease is observed. Data obtained with a second run of measurements for
$Fe_{54}Cr_{46}$ are added, and they agree well with those obtained within the first run.
The critical temperature at which
the anomaly starts, 31.3~K and 16.3~K, for $Fe_{54}Cr_{46}$ and
$Fe_{52}Cr_{48}$, respectively, was determined for each sample from the
intersection of the curved lines, representing the behavior expected from the
Debye model, with the straight lines, representing the anomalous part of the
data.   A good agreement between the temperatures
 at which the anomalies in $\langle CS\rangle $ occur and the corresponding Curie temperatures, can
be taken as evidence that the anomaly in $\langle CS\rangle $, hence in the atomic
vibrations, is related to the magnetic state of the samples.

\begin{figure*}[tp]
\includegraphics[width=.95\textwidth]{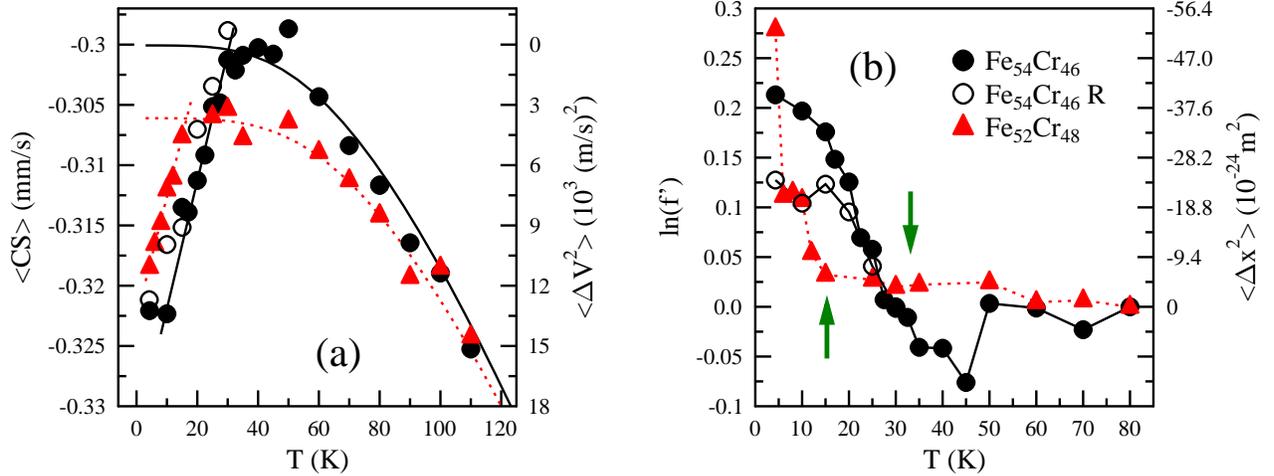}%
\caption{(color online)
(a) The average center shift, $\langle CS\rangle $, versus temperature, $T$, for the
investigated samples: Circles stand for the $Fe_{54}Cr_{46}$ sample (full for the first run and open for the second run),  while the trangles are for the $Fe_{52}Cr_{48}$. The solid and the dashed curved lines represent the behavior
expected from the Debye model.  Their intersection with the oblique straight lines marks
the temperature at which the anomaly in $\langle CS\rangle $ starts to occur. The right-hand axis is scaled with the corresponding change of the mean-square velocity calculated from equ. (2),
(b) ln$f'$, $f'$ being a measure for the relative recoil-free fraction, versus
temperature, $T$, for the investigated samples.  The Curie temperatures, as obtained for the two samples,
are indicated by arrows.
}
\label{F02}
\end{figure*}

A further support to this supposition can be lend from the spectra measured at 4.2 K in an external magnetic field, $B_o$ - see Fig. 3a.
These spectra were analyzed in terms of a standard hyperfine field distribution method, assuming a linear correlation between the hyperfine field and the isomer shift.
The values of $\langle CS\rangle $ derived from this approach are displayed in Fig. 3b versus the external magnetic field, $B_o$,
showing a significant dependence on $B_o$. The increase of the amplitude of $\langle CS\rangle $ with $B_o$ is equivalent to the increase
 of the mean-square velocity. The effect is consistent with the results found from the zero-field spectra - Fig. 2a.

\begin{figure*}[tp]
\includegraphics[width=.95\textwidth]{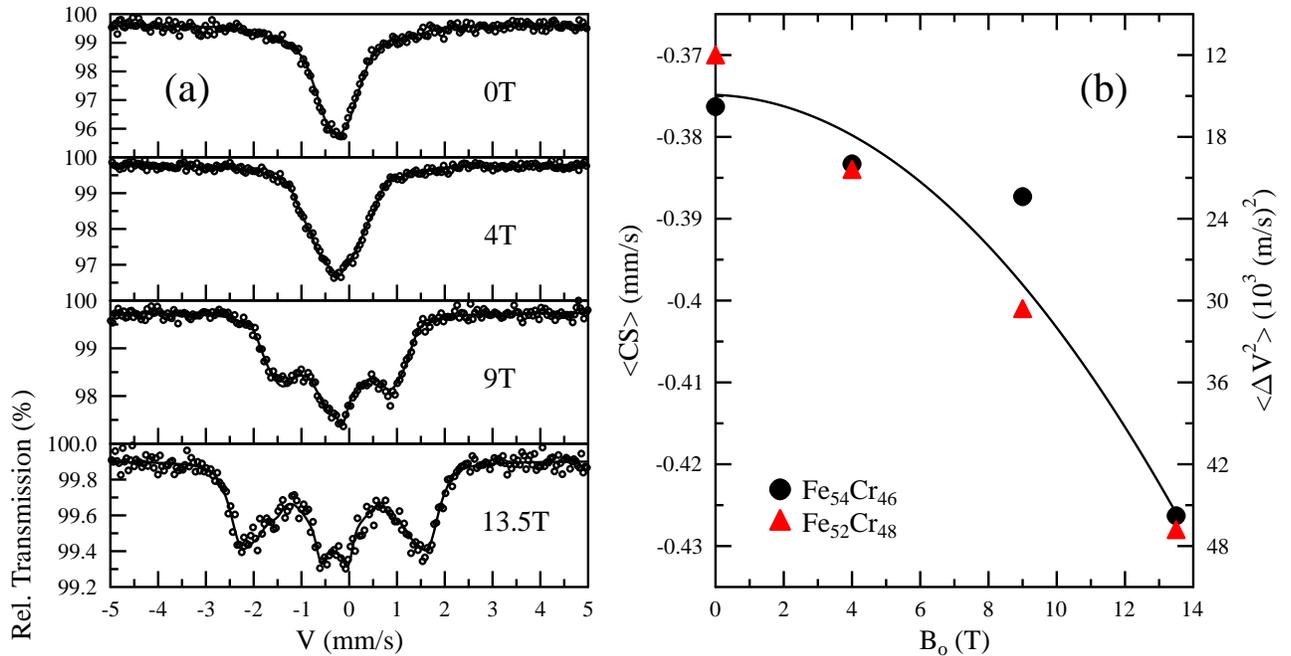}%
\caption{(color online)
(a) $^{57}$Fe M\"ossbauer spectra recorded for the $\sigma-Fe_{54}Cr_{46}$ samples in an external
magnetic field, $B_o$, and
(b) The average center shift, $\langle CS\rangle $, versus $B_o$ for the
investigated samples.  The solid line is a parabolic fit to the data.
The right-hand axis is scaled in the corresponding change of the mean-square
velocity calculated from equ. (2).}
\label{F03}
\end{figure*}

Concerning now the $f$-factor, in the thin-absorber approximation, which was
the case here, it is proportional to the spectral area, $A$.  The
$ln(f')$, where $f'=A/A_o$, $A_o$ being the spectral area at 80 K, is shown for both samples in Fig.  2b
versus $T$.  The right-hand axis is scaled in the underlying
change of the mean-square amplitude of vibrations, $\Delta\langle x^2\rangle $, relative to its value at 80 K.  One can
easily notice that an increase in $ln(f')$ is observed for both samples below
$T$ close to the corresponding Curie points (indicated by arrows).  The
increase in $f'$ on decreasing $T$ is equivalent to the decrease of the
mean-square amplitude of vibrations, and it indicates a hardening of the lattice.

The observed changes in $\langle v^2\rangle $ and in $\langle x^2\rangle $ can be also expressed in terms of
underlying changes in the kinetic, $\Delta E_k$, and in  the potential, $\Delta E_p$, energies of the vibrating
atoms.  A change in the kinetic energy with respect to its value at $T_c$ for the
$Fe_{54}Cr_{46}$ sample is presented in Fig.  4a.  It is evident
that below the Curie point, $T_c$, the kinetic energy increases reaching its maximum of ca.  4
meV  at 4.2 K.  Such a "non-thermodynamic" behavior could be related to
a spin-phonon coupling which sets in on entering the magnetic state and it becomes stronger as $T$ decreases.

The change of the potential
energy in the harmonic approximation, $\Delta E_p=0.5D\Delta{\langle x^2\rangle }$, can be evaluated using  for $D$ (a spring
constant) the value of 155 N/m as found elsewhere \cite{Dubiel10}.  The
results obtained for the $Fe_{54}Cr_{46}$ sample are shown in Fig.  4b.  It
can be seen that $\Delta E_p$ show an opposite trend than $\Delta E_k$ does, as it starts to decrease at $T$ close to $T_c$.
 At $T$ equal to about 15 K, $\Delta E_p$ reaches the value of about 2 meV which hardly
depends on $T$ for its lower values.

Knowing  changes in both forms of the mechanical energy, one can calculate a change of the
total mechanical energy of the atomic vibrations, $\Delta E=\Delta E_k + \Delta E_p$, for the temperature
range of interest.  The behavior of $\Delta E$, as presented in Fig. 4c,  resembles that of
$\Delta E_p$ i.e.  it is constant below ca.  15 K, and increases steeply in the
range of $\sim 15$ K $< T < \sim 33$ K.  This kind of behavior follows from the fact that the dominant
contribution to the observed anomaly as expressed in terms of energy is due
to the potential energy.  In other words, the effect of magnetism on the
atomic vibrations in the studied samples manifest itself mainly via the
decrease of the mean-square amplitude of vibrations, and, to a much less
extent, through the accompanying increase of the mean-square velocity of
these vibrations.  The lack of balance in the behavior of the kinetic and the
potential energy means that the vibrations are not harmonic i.e. they cannot be described properly
in terms of the Debye model.  However, both
behaviors, could be interpreted as indicative of a hardening of the lattice
vibrations caused by a magnetic state.  Such behavior is, to our best
knowledge, unique as previously observed anomalies were different.  In
particular, \cite{Shechter76} using the same technique (MS)
for $DyFe_2$ a similar to ours anomaly in $CS$ but an opposite one in $f$ was observed.
However, in that case, the material was strongly magnetostrictive, hence the
observed anomalies must not necessarily be caused by the spin-phonon
coupling.  In our case, no traces of the magnetostriction were observed
(lattice constants did not show any anomaly, and below 100 K they were hardly
dependent on $T$), hence the observed anomalies and departure from the Debye-like dynamics
 seem to be directly related to the spin-phonon coupling.

\begin{figure}[t]
\includegraphics[width=.49\textwidth]{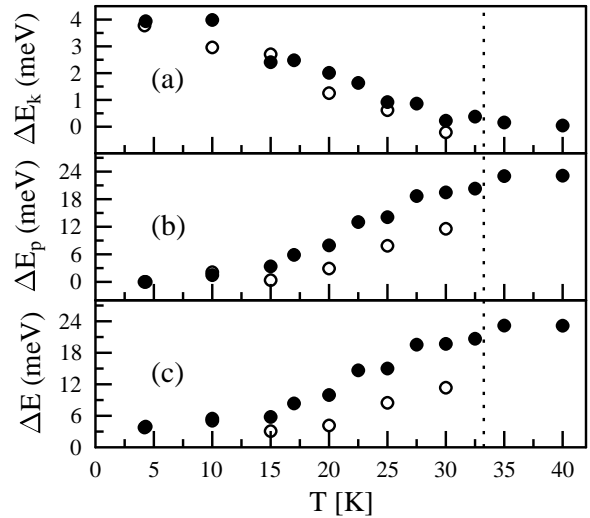}%
\caption{
Changes in the (a) kinetic, $\Delta E_k$, (b) potential, $\Delta E_p$ and (c) total mechanical energy, $\Delta E$
versus temperature, $T$, as found for the $Fe_{54}Cr_{46}$ sample in the temperature range where the anomaly
was observed. Full circles are for the first run and open for the second one. The dashed vertical line indicate the Curie point.
}
\label{F04}
\end{figure}

 In summary, we have revealed that both
spectral quantities viz.  the centre shift, $CS$, and the recoil-free
fraction, $f$, exhibit a strong anomaly on entering the magnetic state in the
studied samples.  They both indicate a hardening of the lattice, which was to
our best knowledge not observed so far.  Anomalous changes in $CS$ and in $f$
were expressed in the underlying changes in the kinetic and in the potential
energy of atomic vibrations, respectively, revealing a very unusual (non-thermodynamic) behavior
of the former viz.  an increase with lowering temperature.  This and a lack
of balance between the two forms of energy ($E_p > E_k$) can be
understood, if new degrees of freedom were opened below $T_c$.  The
opening might be related to the spin-phonon coupling that becomes operative
when the sample becomes magnetically ordered, causing the observed lack of energy conservation and the non-Debye like vibrations.

\begin{acknowledgments}
The project was carried out within the Austrian-Polish scientific cooperation
(project 16/01).  The Ministry of Science and Higher Education, Warsaw, is
acknowledged for support.  \end{acknowledgments}

\providecommand{\noopsort}[1]{}\providecommand{\singleletter}[1]{#1}%

\end{document}